\documentclass[a4paper]{article}

\usepackage{INTERSPEECH2019}

\title{Multi-Accent Adaptation based on Gate Mechanism}
\name{Han Zhu$^{1,2}$, Li Wang$^{1}$, Pengyuan Zhang$^{1,2}$, Yonghong Yan$^{1,2,3}$}
\address{
  $^1$Key Laboratory of Speech Acoustics and Content Understanding, Institute of Acoustics, China\\
  $^2$University of Chinese Academy of Sciences, China\\
  $^3$Xinjiang Laboratory of Minority Speech and Language Information Processing,
Xinjiang Technical Institute of Physics and Chemistry, Chinese Academy of Sciences, China}
\email{\{zhuhan,wangli,zhangpengyuan,yanyonghong\}@hccl.ioa.ac.cn}

\begin{document}

\maketitle
\begin{abstract}
When only a limited amount of accented speech data is available, to promote multi-accent speech recognition performance, the conventional approach is accent-specific adaptation, which adapts the baseline model to multiple target accents independently. To simplify the adaptation procedure, we explore adapting the baseline model to multiple target accents simultaneously with multi-accent mixed data. Thus, we propose using accent-specific top layer with gate mechanism (AST-G) to realize multi-accent adaptation. Compared with the baseline model and accent-specific adaptation, AST-G achieves 9.8\% and 1.9\% average relative WER reduction respectively. However, in real-world applications, we can't obtain the accent category label for inference in advance. Therefore, we apply using an accent classifier to predict the accent label. To jointly train the acoustic model and the accent classifier, we propose the multi-task learning with gate mechanism (MTL-G). As the accent label prediction could be inaccurate, it performs worse than the accent-specific adaptation. Yet, in comparison with the baseline model, MTL-G achieves 5.1\% average relative WER reduction. 
\end{abstract}

\section{Introduction}

Accent is known to be one of the key factors to degrade the performance of automatic speech recognition (ASR) systems \cite{huang2004accent}. Despite great progress over the last few years, the performance gap between the standard and the accented speech recognition remains large. Accent is mainly caused by non-native speakers or native speakers who speak the dialects of the language. The latter is the primary source of mandarin accents \cite{zheng2005accent}. In real-world applications, we often deal with multi-accent problem because the ASR system would be used by speakers with a diverse set of accents.

There is normally a large amount of standard mandarin speech data, while accented mandarin speech data is relatively less. In order to utilize all these data well, the conventional approach would be to train a baseline model with a large amount of standard mandarin speech data and then adapt the baseline model to a specific accent with respective data to produce the accent dependent model, which we will refer to as accent-specific adaptation.

In terms of adaptation methods, traditional ones for GMM-HMM model, such as MLLR \cite{leggetter1995maximum} and fMLLR \cite{gales1998maximum}, have shown great performance. As deep neural network (DNN) was shown to be more powerful than GMM in acoustic model for large vocabulary continuous speech recognition systems \cite{hinton2012deep}, there were many works conducted on DNN adaptation. Selective fine-tuning \cite{sim2018domain} is a very intuitive method to alleviate the data mismatch problem. Linear transformation, such as linear input network (LIN) \cite{li2010comparison}, linear hidden network (LHN) \cite{gemello2006adaptation} and linear output network (LON) \cite{li2010comparison} were shown to be effective when adaptation data is relatively less. Regularization based methods \cite{liao2013speaker}\cite{yu2013kl} were introduced to solve the overfitting problem during adaptation. Furthermore, cluster adaptive training (CAT) \cite{tan2015cluster}, factorized hidden layer adaptation (FHL) \cite{samarakoon2016factorized}, learning hidden unit contributions (LHUC) \cite{swietojanski2014learning} and speaker code \cite{abdel2013fast} were proposed to translate the original problem into evaluation of a small number of parameters during adaptation.

Accent-specific adaptation is cumbersome considering that multiple accent dependent models will be produced and we have to find the best configuration for them individually. Therefore, an integrated adapted model would be desirable. To that end, we can just adapt the baseline model to multiple target accents simultaneously with multi-accent mixed data in an integrated model, which we will refer to as multi-accent adaptation. Adaptation methods like CAT, FHL, and speaker code require multi-class quantitatively balanced data during the baseline model training procedure, which makes it impossible to transfer these methods to multi-accent adaptation. In terms of other adaptation methods like selective fine-tuning and  linear transformation, we found it empirically that multi-accent adaptation based on them can improve speech recognition performance, but still performs worse than the accent-specific adaptation.

In this literature, we proposed two novel approaches to promote the performance of multi-accent adaptation. Our work concentrated on the neural network part of the acoustic model. We kept the rest of the ASR system, such as HMM and language model, to be unchanged. Firstly, we proposed accent-specific top layer with gate mechanism (AST-G). In AST-G, on the one hand accent-specific top layer \cite{huang2014multi} is used to model the distinctive characteristics of multiple target accents via different parameter sets, on the other hand gate mechanism \cite{kim2018towards} is used to modulate the network’s internal representations in an accent-specific way with the help of accent category labels. Experiments showed that AST-G can achieve 9.8\% and 1.9\% average relative WER reduction compared with the baseline model and accent-specific adaptation respectively. In order to solve one significant shortcoming of AST-G which is that accent category labels are required for inference, we proposed  multi-task learning \cite{Jain2018} with gate mechanism (MTL-G), in which the primary task is context-dependent tied phone state classification and the secondary task is accent classification. The output of the secondary task network is used as the accent category label and fed into the gate unit of the primary task network.

The rest of the paper is organized as follows, in Section \ref{sec:AST-G} and Section \ref{sec:MTL-G}, we describe our proposed approaches in detail. Then, we evaluate our proposed approaches and compare them with commonly used adaptation methods in Section \ref{sec:Experiments}, and conclude the study in Section \ref{sec:Conclusions}.

\section{Accent-specific top layer with gate mechanism (AST-G)}\label{sec:AST-G}
\begin{figure}[t]
  \centering
  \includegraphics[width=\linewidth]{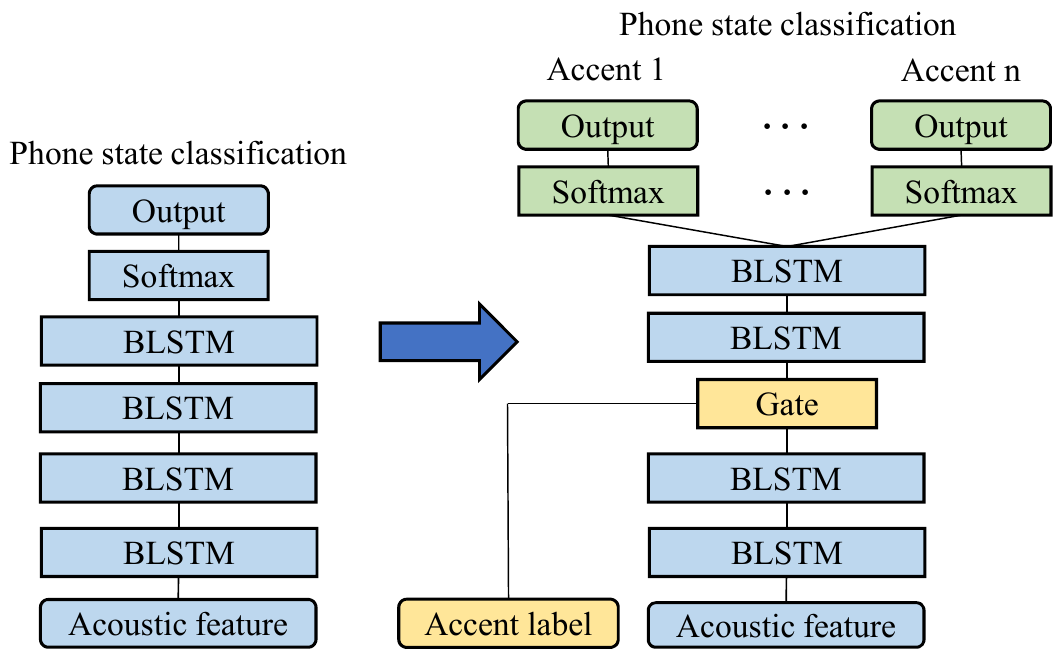}
  \caption{Accent-specific top layer with gate mechanism (AST-G).}
  \label{fig:approach1}
\end{figure}

In this approach, we assume a prior that accent category labels are available for both adaptation and inference. As figure~\ref{fig:approach1} demonstrated, gate unit (yellow) and accent-specific top layers (green) are added on the basis of the baseline model (blue) before adaptation. We use bidirectional long short-term memory (BLSTM) recurrent neural network as the specific DNN in our experiments for its outstanding performance in acoustic model.

\subsection{Gate mechanism}\label{gate}

The speech spoken by speakers with different accents differs in the phonetic pronunciation. Utilizing these differences in acoustic model was proven to be useful \cite{Zheng+2016}. However, using manufactured phonetic information is quite subjective and cumbersome. It would be convenient if we only provide accent category labels and let the neural network model the difference in phonetic information between target accents.  And our gate mechanism is a way to do so. Our work is similar to the one in \cite{kim2018towards} which is conducted on the multilingual ASR task. We use accent category labels to modulate the network’s internal representations in an accent-specific way. To do so, the output of the $i_{th}$ layer $h_i$ and the accent category label (one-hot vector) $v_a$ are fed into the gate unit. After some operations, the gate value $g(h_i,v_a)$ is passed to the $(i+1)_{th}$ layer.
Several concrete realizations of the gate unit can be used. The operations of these different gate units are described below. 
\begin{itemize}
\item Addition (which we will refer to as the GATE \uppercase\expandafter{\romannumeral1}),
\begin{equation}
  g(h_i,v_a) = h_i+Vv_a+b
  \label{eq1}
\end{equation}
where $V$ is a matrix of size $M \times N$ if dimensions of $h_i$ and $v_a$ are $M$ and $N$ respectively, and $b$ is a bias vector.

\item More complex addition (which we will refer to as the GATE \uppercase\expandafter{\romannumeral2}),
\begin{equation}
  g(h_i,v_a) = Uh_i+Vv_a+b
  \label{eq2}
\end{equation}
where $U$ is an identity matrix.

\item Addition with a sigmoid activation function (which we will refer to as the GATE \uppercase\expandafter{\romannumeral3}),
\begin{equation}
  g(h_i,v_a) = \sigma(h_i+Vv_a+b)
  \label{eq3}
\end{equation}
Where $\sigma()$ denotes the sigmoid function.

\item	Elementwise multiplication (which we will refer to as the GATE \uppercase\expandafter{\romannumeral4}),
\begin{equation}
  g(h_i,v_a )=h_i \cdot Vv_a+b
  \label{eq4}
\end{equation}

\item	Addition and elementwise multiplication (which we will refer to as the GATE \uppercase\expandafter{\romannumeral5}),
\begin{equation}
  g(h_i,v_a )=h_i \cdot (h_i+Vv_a+b)
  \label{eq5}
\end{equation}
\end{itemize}

\subsection{Accent-specific top layer}

The accent-specific top layer \cite{huang2014multi} was motivated by the shared hidden layers of deep neural network in the multilingual speech recognition \cite{huang2013cross}. It can be seen as a similar way to gate mechanism to model the distinctive characteristics of multiple target accents. Before adaptation, the top layer of the baseline model is duplicated to build multiple accent-specific top layers. During adaptation, in our approach, accent-specific top layers are updated together with the extra parts of the network, which is different from the original one which only update the parameters of  accent-specific top layers \cite{huang2014multi}. The accent-specific top layer will only be updated with the specific target accented speech data, while the shared bottom layers will be updated with all target accented speech data. After adaptation, the inference of a specific target accented speech will be done using the respective accent-specific top layer and the shared bottom layers.

The shared accent independent bottom layers allow maximal knowledge transfer between multiple target accents. This is especially important when only small amount of accented speech data is available \cite{huang2014multi}. 

\subsection{AST-G}
AST-G uses the above two methods to further improve the performance, which make sense because they have the same goal to model multiple target accents accurately by using either accent category labels or accent-specific parameter sets.

\section{Multi-task learning with gate mechanism (MTL-G)}\label{sec:MTL-G}
\begin{figure}[t]
  \centering
  \includegraphics[width=\linewidth]{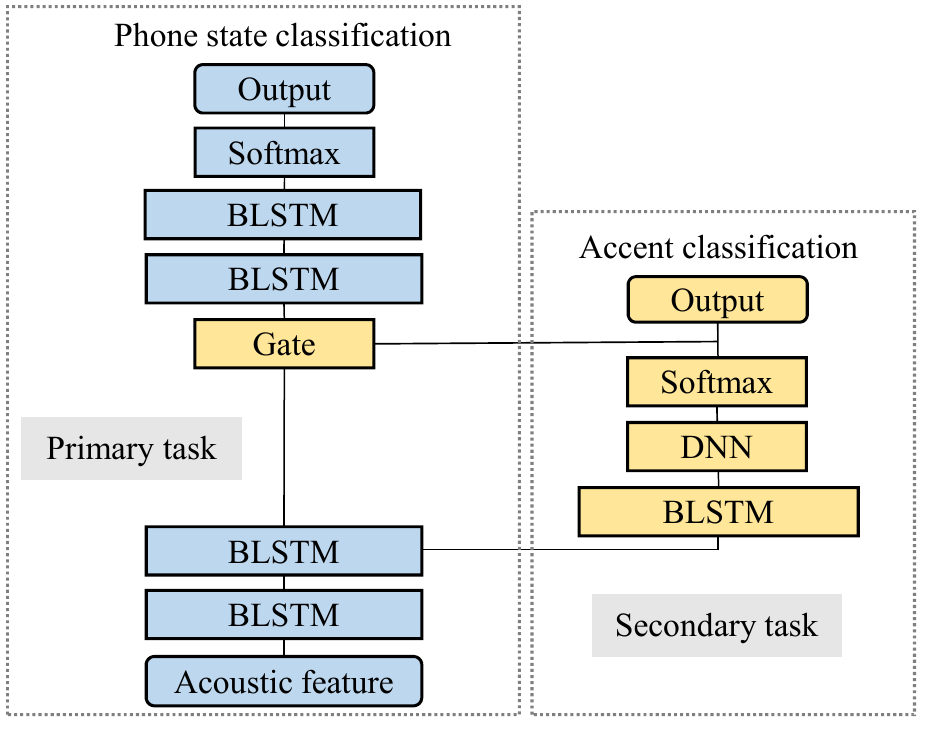}
  \caption{Multi-task learning with gate mechanism (MTL-G).}
  \label{fig:approach2}
\end{figure}

This approach is demonstrated in Figure~\ref{fig:approach2}. The network jointly predicts context-dependent tied phone states (the primary task) and accent category labels (the secondary task). Unlike AST-G, the extra information fed into the gate unit is the output of the secondary task network, which can be seen as the soft accent category label. Therefore, we only need to know the accent category of the utterances in the adaptation procedure. Before adaptation, the gate unit and the secondary task network (yellow) are added on the basis of the baseline model (blue).

During adaptation, the whole network is trained using backpropagation with the following loss function,
\begin{equation}
  L_{total}=(1-\lambda)L_{primary}+\lambda L_{secondary}   
  \label{eq6}
\end{equation}
where $\lambda$ is a weight hyperparameter that is used to linearly interpolate the individual loss terms.

After adaptation, the entire network will predict the tied phone states together. 

The promising aspect of MTL-G is that the accent category label won’t be required for inference, which would be convenient in real-world applications.

\section{Experiments and Results}\label{sec:Experiments}
\subsection{Data and baseline model description}

In our experiments, training set for the baseline model consists of 7000 hours of standard mandarin speech data. Adaptation set consists of 4 kinds of accents which are Anhui (AH), Beijing (BJ), Shanxi (SX) and Yunnan (YN) and each accent has 20 hours of data. Evaluation set consists of above 4 accents and additional 3 unseen accents which are Guizhou (GZ), Sichuan (SC) and Zhejiang (ZJ). Each accent in evaluation set has 2 hours of data. All the data is conversational telephone speech in real-world customer service scenario. 60-dimensional PLP features are used in the experiments.

All our ASR systems were implemented using the Kaldi toolkit \cite{povey2011kaldi}. Our baseline model to be adapted has 4 BLSTM layers and each layer has 512 hidden units for each direction. The targets are 5000 context-dependent tied phone states.

\subsection{Comparison of selective fine-tuning and linear transformation}

We first evaluated the performances of multi-accent adaptation using selective fine-tuning and linear transformation (LIN, LON and LHN).

For selective fine-tuning, we update the parameters of the first $n$ BLSTM layers, denoted as \bfseries fine-tune \mdseries$\bm{(n)}$. The structure of our linear transformation follows the one in \cite{kitza2018comparison}. \bfseries LIN \mdseries is a linear transformation layer inserted after the input layer. LHN consists of two independent linear transformation layers inserted after the forward and backward direction of the $i_{th}$ hidden layer, denoted as \bfseries LHN \mdseries $\bm{i}$. \bfseries LON \mdseries is a linear transformation layer inserted between the weight matrix of output layer and the softmax of the network. Weight matrixes of these linear transformation layers are initialized to the identity matrix and the bias vectors are initialized to zero. 

\begin{table}[th]
  \caption{WERs (\%) of models using selective fine-tuning and linear transformation }
  \label{tab:table1}
  \centering
  \begin{tabular}{ c@{} c@{}c@{} c@{}c@{} c@{}}
    \toprule
    \multicolumn{1}{c}{\textbf{model}} & \multicolumn{1}{c}{\textbf{AH}} & \multicolumn{1}{c}{\textbf{BJ}} & \multicolumn{1}{c}{\textbf{SX}} & \multicolumn{1}{c}{\textbf{YN}} & \multicolumn{1}{c}{\textbf{AVE}}  \\
    \midrule
    baseline                 &  28.9  & 9.8  & 30.9 & 24.3 & 23.5            \\
    fine-tune(1)   & 28.4  & 10.2  & 30.3 & 24.1 & 23.3               \\
    fine-tune(2)   & 28.0  & 10.0  & 29.9 & 23.9 & 23.0     \\
    fine-tune(3)   & 27.6  & 9.9  & 29.6 & 23.7 & 22.7             \\
    fine-tune(4)   & 27.6  & 9.9  & 29.5 & 23.8 & 22.7             \\
    fine-tune(all)          & 27.5  & 9.9  & 29.4 & 23.7 & \bfseries 22.6 \mdseries \\
    LIN                        & 28.7	& 10.2 & 30.3	& 24.2	& 23.4          \\
    LHN 1                     &  28.4	& 10.2	& 30.5	& 24.4	& 23.4             \\
    LHN 2                     & 28.4	& 10.6	& 30.1	& 24.2	& 23.3             \\
    LHN 3                     &  28.0	& 10.3	& 30.0	& 24.0	& 23.1             \\
    LHN 4                     &  28.1& 	9.7	& 30.0	& 24.2	&\bfseries  23.0 \mdseries             \\
    LON                      &  28.1	& 10.2	& 30.1	& 24.4	& 23.2           \\
    accent-specific     & 26.4	& 8.8	& 28.7	& 22.5	& \bfseries 21.6 \mdseries            \\
    \bottomrule
  \end{tabular}
  
\end{table}

WERs (word error rate) of 4 accents are shown in Table~\ref{tab:table1}. \bfseries AVE \mdseries denotes average WER of all 4 accent evaluation sets. We mainly focus on the average WER of 4 accents, which roughly represents the performance of the system in the multi-accent applications. It turns out that multi-accent adaptation using fine-tuning all layers can get best performance. The performance on every accent except Beijing improves more or less. In terms of linear transformation, we find that LHN inserted after the last hidden layer (LHN 4) is optimal, but is still worse than fine-tuning all layers. The reason why the linear transformation based methods don't work well is that the data used in the adaptation procedure is relatively more than the conventional problem the linear transformation is used to solve, such as speaker adaptation within a few utterances. With 80 hours of data, fine-tuning the entire model is relatively reliable and can adapt the model to multiple targets accents better. 

In addition, we used the best method, fine-tuning all layers, to realize accent-specific adaptation. The hyperparameters were optimized for each model. The results are shown in the last row in Table~\ref{tab:table1}. We notice that this method achieves 1\% average absolute WER reduction compared with the multi-accent adaptation using fine-tuning all layers.

\subsection{Improvements using gate unit}

We next evaluated the performances of multi-accent adaptation using different gate units described in section \ref{gate}. All gate units are inserted after the first hidden layer. Initializing all parameters in the gate unit randomly is shown to be optimal. Giving the parameters of the gate unit a bigger learning rate than the rest of network is also beneficial for the performance. 

\begin{table}[th]
  \caption{WERs (\%) of models using different gate units }
  \label{tab:table2}
  \centering
  \begin{tabular}{ c@{} c@{}c@{} c@{}c@{} c@{}}
    \toprule
    \multicolumn{1}{c}{\textbf{model}} & \multicolumn{1}{c}{\textbf{AH}} & \multicolumn{1}{c}{\textbf{BJ}} & \multicolumn{1}{c}{\textbf{SX}} & \multicolumn{1}{c}{\textbf{YN}} & \multicolumn{1}{c}{\textbf{AVE}}  \\
    \midrule
    baseline      &  28.9  & 9.8  & 30.9 & 24.3 & 23.5            \\
    GATE \uppercase\expandafter{\romannumeral1}   & 27.1	& 9.0	& 28.7	& 22.9	& \bfseries21.9  \mdseries                         \\
    GATE \uppercase\expandafter{\romannumeral2}   & 27.7	& 9.2	& 29.6	& 23.8	& 22.6     \\
    GATE \uppercase\expandafter{\romannumeral3}   & 28.1	& 10.4	& 30.1	& 24.7	& 23.3 \\
    GATE \uppercase\expandafter{\romannumeral4}   & 27.0	& 9.2	& 29.3	& 23.6	& 22.3             \\
    GATE \uppercase\expandafter{\romannumeral5}   & 28.4	& 9.9	& 30.4	& 24.3	& 23.3  \\
    \bottomrule
  \end{tabular}
\end{table}

Table~\ref{tab:table2} gives the performances of different gate units. It turns out that the simplest gate GATE \uppercase\expandafter{\romannumeral1} performs best, which is contrast with the conclusion in \cite{kim2018towards}. The main reason is that in \cite{kim2018towards}, the network with gate unit is trained form scratch with relatively more data so that the complex gate unit could be utilized well. However, in our approach, we add the gate unit on the basis of a well-trained baseline model. A complex gate unit would make the adaptation with limited data difficult to get a good performance. 

Then we evaluated how many layers the gate unit applied to is best. We adjust the number of gate units by inserting them after the first $n$ hidden layers, denoted as \bfseries GATE \uppercase\expandafter{\romannumeral1} \mdseries$\bm{(n)}$. The results are shown in Table~\ref{tab:table3}. Inserting gate units after the first 3 layers is best in this experiment.

\begin{table}[th]
  \caption{WERs (\%) of models using different number of gate units }
  \label{tab:table3}
  \centering
  \begin{tabular}{ c@{} c@{}c@{} c@{}c@{} c@{}}
    \toprule
    \multicolumn{1}{c}{\textbf{model}} & \multicolumn{1}{c}{\textbf{AH}} & \multicolumn{1}{c}{\textbf{BJ}} & \multicolumn{1}{c}{\textbf{SX}} & \multicolumn{1}{c}{\textbf{YN}} & \multicolumn{1}{c}{\textbf{AVE}}  \\
    \midrule
    baseline      &  28.9  & 9.8  & 30.9 & 24.3 & 23.5            \\
    GATE \uppercase\expandafter{\romannumeral1} (1)   & 27.1	& 9.0	& 28.7	& 22.9	& 21.9                       \\
    GATE \uppercase\expandafter{\romannumeral1} (2)   & 26.9	& 9.0	& 28.8	& 22.6	& 21.8     \\
    GATE \uppercase\expandafter{\romannumeral1} (3)   & 26.6	& 8.8	& 28.4	& 22.4	  & \bfseries 21.6 \mdseries    \\
    GATE \uppercase\expandafter{\romannumeral1} (4)   & 26.6	& 8.9	& 28.5	& 22.6	& 21.7           \\
    \bottomrule
  \end{tabular}

\end{table}

\subsection{Improvements using accent-specific top layer}
Furthermore, we evaluated the performance of multi-accent adaptation using top layer. A bigger learning rate is adopted for the top layer, which make sense for that less data is fed into the accent-specific top layer than the bottom layers. We use learning rate factor (LR factor) to denote the ratio of the learning rate for accent-specific top layer to the one for bottom layers \cite{ghahremani2017investigation}. The performances of different LR factors are shown in Table~\ref{tab:table4}. The best LR factor is shown to be 10. 

\begin{table}[th]
  \caption{WERs (\%) of models using accent-specific top layer with different learning rate factors }
  \label{tab:table4}
  \centering
  \begin{tabular}{ c@{} c@{}c@{} c@{}c@{} c@{}}
    \toprule
    \multicolumn{1}{c}{\textbf{model}} & \multicolumn{1}{c}{\textbf{AH}} & \multicolumn{1}{c}{\textbf{BJ}} & \multicolumn{1}{c}{\textbf{SX}} & \multicolumn{1}{c}{\textbf{YN}} & \multicolumn{1}{c}{\textbf{AVE}}  \\
    \midrule
    baseline      &  28.9  & 9.8  & 30.9 & 24.3 & 23.5            \\
    LR factor = 2   & 26.9	& 9.7	& 29.1	& 23.4	& 22.3         \\
    LR factor = 5   & 26.8	& 9.6	& 29.0	& 23.2	& 22.2     \\
    LR factor = 10   & 26.6	& 9.5	& 28.9	& 23.0	  & \bfseries 22.0 \mdseries    \\
    LR factor = 20   & 26.6	& 9.4	& 29.5	& 23.3	& 22.2         \\
    \bottomrule
  \end{tabular}
 
\end{table}

\subsection{Improvements using accent-specific top layer with gate mechanism (AST-G)}

Eventually we evaluated our proposed AST-G approach. We used the best configuration of the above two methods in this experiment. The results are shown in Table~\ref{tab:table5}. The model using AST-G performs better than accent-specific adaptation on every target accent.

\begin{table}[th]
  \caption{WERs (\%) of the model using AST-G }
  \label{tab:table5}
  \centering
  \begin{tabular}{ c@{} c@{}c@{} c@{}c@{} c@{}}
    \toprule
    \multicolumn{1}{c}{\textbf{model}} & \multicolumn{1}{c}{\textbf{AH}} & \multicolumn{1}{c}{\textbf{BJ}} & \multicolumn{1}{c}{\textbf{SX}} & \multicolumn{1}{c}{\textbf{YN}} & \multicolumn{1}{c}{\textbf{AVE}}  \\
    \midrule
    baseline      &  28.9  & 9.8  & 30.9 & 24.3 & 23.5            \\
    fine-tune(all)          & 27.5  & 9.9  & 29.4 & 23.7 & 22.6  \\
    accent-specific     & 26.4	& 8.8	& 28.7	& 22.5	&  21.6             \\
    \bfseries AST-G \mdseries   &\bfseries 26.1 \mdseries	&\bfseries 8.7 \mdseries	&\bfseries 28.0 \mdseries	&\bfseries 22.1 \mdseries	& \bfseries 21.2 \mdseries  \\
    \bottomrule
  \end{tabular}
 
\end{table}

\subsection{Improvements using multi-task learning with gate mechanism (MTL-G)}

In this experiment, MTL-G uses 1 BLSTM layer and 1 DNN layer with sigmoid activation function at the top of the shared bottom BLSTM layers as the secondary task network. The dimensionalities of these 2 layers are 512 and 256 respectively. The best gate unit GATE \uppercase\expandafter{\romannumeral1} is adopted. During adaptation, the shared bottom layers will be only updated using the gradient computed from the primary task network, which aims to avoid the disturbance from the secondary task network. The results are listed in Table~\ref{tab:table6}. We find that MTL-G performs better than multi-accent adaptation using fine-tuning but worse than accent-specific adaptation. The performance degradation is tolerable because accent category labels won't be required for inference and it is very convenient in real-world applications.

\begin{table}[th]
  \caption{WERs (\%) of the model using MTL-G. }
  \label{tab:table6}
  \centering
  \begin{tabular}{ c@{} c@{}c@{} c@{}c@{} c@{}}
    \toprule
    \multicolumn{1}{c}{\textbf{model}} & \multicolumn{1}{c}{\textbf{AH}} & \multicolumn{1}{c}{\textbf{BJ}} & \multicolumn{1}{c}{\textbf{SX}} & \multicolumn{1}{c}{\textbf{YN}} & \multicolumn{1}{c}{\textbf{AVE}}  \\
    \midrule
    baseline      &  28.9  & 9.8  & 30.9 & 24.3 & 23.5            \\
    fine-tune(all)          & 27.5  & 9.9  & 29.4 & 23.7 & 22.6  \\
    accent-specific     & 26.4	& 8.8	& 28.7	& 22.5	&  21.6             \\
    \bfseries MTL-G \mdseries   &\bfseries 27.3 \mdseries	&\bfseries 9.7  \mdseries	& \bfseries 28.9	 \mdseries& \bfseries 23.4	  \mdseries& \bfseries 22.3 \mdseries  \\
    \bottomrule
  \end{tabular}
 
\end{table}

   Furthermore, we tested on 3 unseen accents consisting of Guizhou (GZ), Sichuan (SC) and Zhejiang (ZJ). The results are shown in Table~\ref{tab:table7}. We surprisingly find that this approach can perform well on the unseen accents. 

\begin{table}[th]
  \caption{WERs (\%) of the model using MTL-G on 3 unseen accents }
  \label{tab:table7}
  \centering
  \begin{tabular}{ c@{} c@{} c@{} c@{}}
    \toprule
    \multicolumn{1}{c}{\textbf{model}} & \multicolumn{1}{c}{\textbf{GZ}}  & \multicolumn{1}{c}{\textbf{SC}}  & \multicolumn{1}{c}{\textbf{ZJ}}   \\
    \midrule
    baseline      &  30.6  &  39.6&  20.9  \\
    \bfseries MTL-G \mdseries   & \bfseries 29.9 \mdseries	 & \bfseries 37.4 \mdseries & \bfseries 20.2 \mdseries  \\
    \bottomrule
  \end{tabular}
 
\end{table}

The performance improvements on target accents and unseen accents show that the multi-accent adaptation using MTL-G could lead to an integrated model which could perform well in multi-accent applications without the requirement of accent category labels for inference.

\section{Conclusions}\label{sec:Conclusions}
In this work, we proposed two multi-accent adaptation approaches to promote the performance of multi-accent speech recognition. AST-G uses accent-specific top layer and gate mechanism to model the distinctive characteristics of multiple target accents. This approach achieves 9.8\% average relative WER reduction compared with the baseline model. Furthermore, to avoid the requirement of accent category labels,  MTL-G jointly learns a multi-accent acoustic model and an accent classifier. In comparison with the baseline model, MTL-G achieves 5.1\% average relative WER reduction.

For future work, we plan to extend MTL-G for its convenience in real-world applications.

\section{Acknowledgements}
This work is partially supported by the National Key Research and Development Program (Nos. 2016YFB0801203, 2016YFB0801200), the National Natural Science Foundation of China (Nos. 11590774, 11590770), the Key Science and Technology Project of the Xinjiang Uygur Autonomous Region (No.2016A03007-1), the Pre-research Project for Equipment of General Information System (No.JZX2017-0994/Y306).

 \bibliographystyle{IEEEtran}

 \bibliography{mybib}


\end{document}